\documentclass[prd,showkeys,floatfix,twocolumn,amsmath,amssymb,floatfix]{revtex4}
\usepackage{graphicx}
\usepackage{epstopdf}
\usepackage{enumitem}
\usepackage{multirow}
\usepackage{subfigure}
\usepackage{bm}
\usepackage{color}
\usepackage[colorlinks,citecolor=blue,anchorcolor=red,menucolor=red,linkcolor=red,filecolor=red,runcolor=red,urlcolor=blue,frenchlinks=red]{hyperref}

\allowdisplaybreaks[3]

\begin{document}

\title{Quantum numbers of excited $\Xi_c^\prime$ and $\Omega_c$ baryons and the $P$-wave $\Sigma_c$ spectrum}

\author{Xuan Luo$^1$}\thanks{These authors contributed equally to this work.}
\author{Shu-Wei Zhang$^1$}\thanks{These authors contributed equally to this work.}
\author{Hua-Xing Chen$^1$}
\email{hxchen@seu.edu.cn}
\author{Hui-Min Yang$^2$}
\email{hmyang@pku.edu.cn}
\affiliation{
$^1$School of Physics, Southeast University, Nanjing 210094, China
\\
$^2$School of Physics, Henan Normal University, Henan 453007, China
}

\begin{abstract}
Heavy quark effective theory (HQET) permits seven $P$-wave states in each of the $\Sigma_c$, $\Xi_c^\prime$, and $\Omega_c$ flavor-sextet sectors, yet the experimentally visible spectra differ strikingly among the three sectors. We show that strong decays, kinematic thresholds, and configuration mixing act as a dynamical filter between the underlying multiplet structure and the observed resonance spectrum. Using QCD sum rules for the masses and light-cone sum rules for the strong decays, we find that only four $\Sigma_c$, four $\Xi_c^\prime$, and five $\Omega_c$ states are sufficiently narrow to contribute to experimentally identifiable resonance structures. The four narrow $\Xi_c$ states, $\Xi_c(2882)$, $\Xi_c(2923)$, $\Xi_c(2939)$, and $\Xi_c(2965)$, and the four narrow $\Omega_c$ states, $\Omega_c(3000)$, $\Omega_c(3050)$, $\Omega_c(3066)$, and $\Omega_c(3090)$, exhibit similar mass-splitting patterns and provide spectroscopic anchors for a common $\lambda$-mode multiplet structure, supporting the successive assignments $J^P=1/2^-,3/2^-,3/2^-$, and $5/2^-$. We tentatively interpret $\Omega_c(3119)$ as a predominantly $\rho$-mode state with $J^P=3/2^-$. With this multiplet structure constrained by the $\Xi_c$ and $\Omega_c$ spectra, we predict four relatively resolvable $P$-wave $\Sigma_c$ states that may overlap within the observed $\Sigma_c(2800)$ and $\Sigma_c(2900)$ structures, providing concrete targets for future amplitude analyses.
\end{abstract}
\pagenumbering{arabic}

\maketitle

$\\$
{\it Introduction}.---
Heavy baryons provide an important laboratory for investigating
nonperturbative quantum chromodynamics and the interplay between
heavy-quark symmetry and the dynamics of the light degrees of
freedom~\cite{Korner:1994nh,Manohar:2000dt,Neubert:1993mb,
Cheng:2015iom,Chen:2016spr,Chen:2022asf,Cheng:2021qpd,
Crede:2024hur}. In the heavy-quark limit, the spin of the heavy
quark decouples from the light subsystem, so that singly heavy
baryons can be systematically organized according to the total
angular momentum and internal configuration of their light
degrees of freedom. For the $P$-wave charmed baryons in the
$SU(3)$ flavor-sextet representation, heavy quark effective
theory permits seven states in each of the $\Sigma_c$,
$\Xi_c^\prime$, and $\Omega_c$ sectors
\cite{Chen:2015kpa,Yang:2021lce}. Establishing how these
underlying heavy-quark multiplets emerge as experimentally
observable resonances is therefore a central problem in
heavy-hadron spectroscopy.

Experimentally, however, the three flavor sectors exhibit
strikingly different spectra. In the $\Sigma_c$ sector, two
structures have been observed in the $\Lambda_c\pi$
invariant-mass spectra,
\begin{equation}
\Sigma_c(2800),\qquad
\Sigma_c(2900),
\end{equation}
by the Belle and LHCb Collaborations
\cite{Belle:2004zjl,LHCb:2026nzu}. In contrast, four relatively
narrow $\Xi_c$ resonances have been reported in the
$\Lambda_c^+K^-$ invariant-mass spectrum
\cite{LHCb:2020iby,LHCb:2022vns}:
\begin{equation}
\Xi_c(2882),\,
\Xi_c(2923),\,
\Xi_c(2939),\,
\Xi_c(2965).
\label{eq:xic_states}
\end{equation}
Five narrow $\Omega_c$ resonances have also been observed in the
$\Xi_c^+K^-$ invariant-mass spectrum
\cite{LHCb:2017uwr,LHCb:2021ptx,LHCb:2023sxp}:
\begin{equation}
\Omega_c(3000),\,
\Omega_c(3050),\,
\Omega_c(3066),\,
\Omega_c(3090),\,
\Omega_c(3119).
\label{eq:omega_states}
\end{equation}
The spin-parity quantum numbers and internal configurations of
many of these states have not yet been conclusively established
\cite{PDG2024,Cheng:2021qpd,Crede:2024hur}. More fundamentally,
it remains unclear why the seven theoretically allowed states in
each flavor sector manifest experimentally as only two
$\Sigma_c$ structures, four relatively narrow $\Xi_c$
resonances, and five narrow $\Omega_c$ resonances.

\begin{figure*}[hbtp]
\begin{center}
\includegraphics[width=0.9\textwidth]{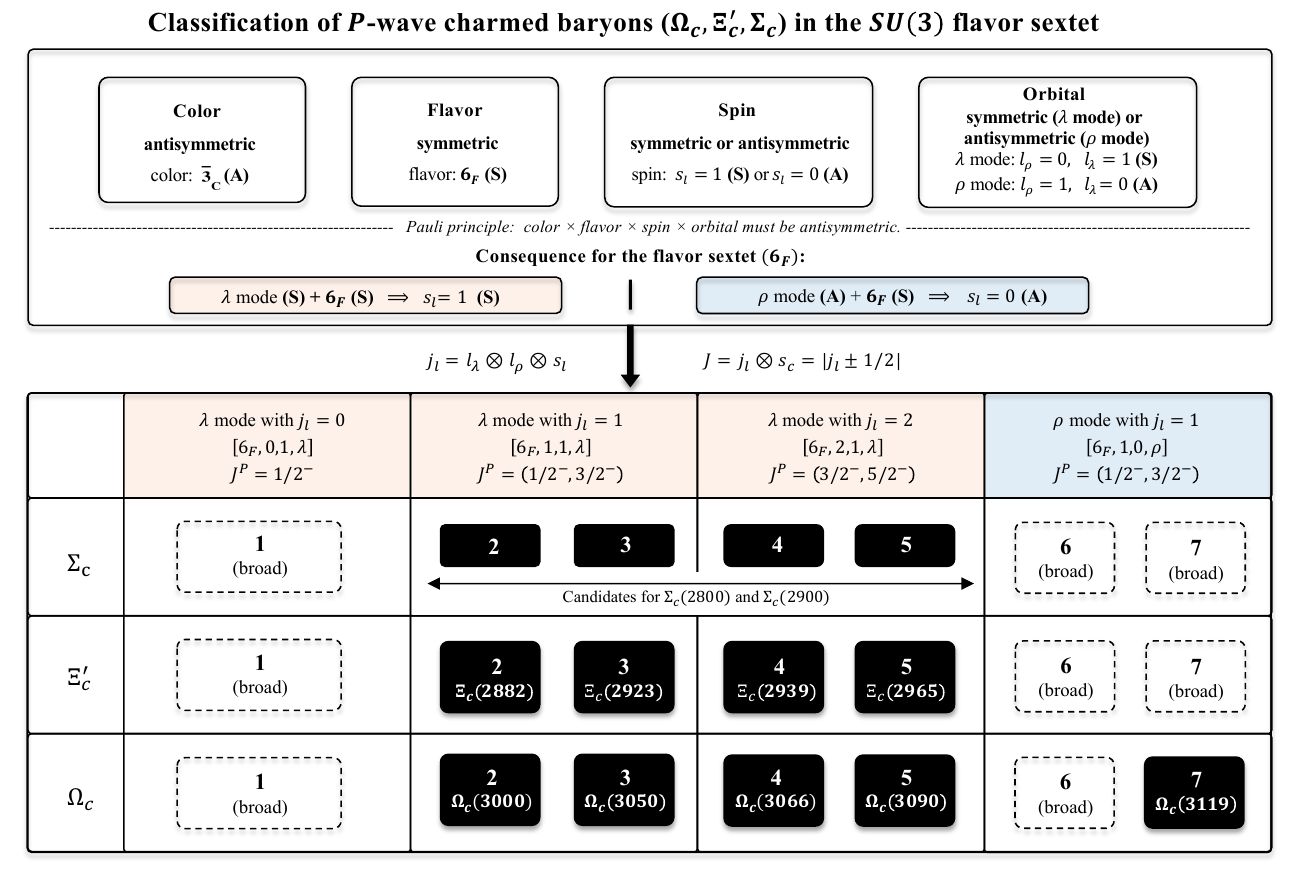}
\end{center}
\caption{
Classification of the $P$-wave charmed baryons in the
$SU(3)$ flavor sextet. The exchange symmetries of the two light
quarks lead to four HQET multiplets and seven states in each of
the $\Sigma_c$, $\Xi_c^\prime$, and $\Omega_c$ sectors. Solid
and dashed boxes denote relatively narrow and broad states,
respectively. The four relatively narrow $\Xi_c^\prime$ states
and five narrow $\Omega_c$ states are assigned to the observed
resonances, while the four resolvable $\Sigma_c$ states may
overlap within the observed $\Sigma_c(2800)$ and
$\Sigma_c(2900)$ structures.
}
\label{fig:classification}
\end{figure*}

Remarkably, the four $\Xi_c$ states and the first four
$\Omega_c$ states listed in Eqs.~\eqref{eq:xic_states} and
\eqref{eq:omega_states} exhibit closely correlated level structures.
Labeling the four states in each sector in order of increasing
mass and defining the three consecutive mass splittings as
$\Delta_i=M_{i+1}-M_i$, we obtain
\begin{equation}
\begin{array}{c@{\qquad}ccc}
\hline
 & \Delta_1 & \Delta_2 & \Delta_3
\\[1mm] \hline
\Xi_c
& 41~{\rm MeV} & 16~{\rm MeV} & 26~{\rm MeV}
\\
\Omega_c
& 50~{\rm MeV} & 16~{\rm MeV} & 24~{\rm MeV}
\\ \hline
\end{array}
\label{eq:intro_splittings}
\end{equation}
The parallel level ordering and similar consecutive splittings
strongly suggest that these states share a common heavy-quark
multiplet organization. The precisely measured narrow states
therefore provide spectroscopic anchors for reconstructing the
underlying heavy-quark multiplets. Once their level ordering and
mixing pattern are constrained, the same structure can be used
to investigate broad and overlapping states that are otherwise
difficult to resolve experimentally.

The excited charmed baryons have been extensively investigated
using quark models
\cite{Copley:1979wj,Ebert:2011kk,Yoshida:2015tia,
Wang:2017kfr,Lu:2020ivo,Weng:2024roa},
lattice QCD
\cite{Padmanath:2013bla,Padmanath:2017lng,Bahtiyar:2020uuj},
heavy-hadron chiral perturbation theory and other effective
approaches
\cite{Cheng:2015naa,Chen:2017gnu,Zhao:2017fov,Ye:2017yvl},
QCD sum rules and light-cone sum rules
\cite{Chen:2015kpa,Chen:2017sci,Wang:2017zjw,
Yang:2020zjl,Yang:2021lce},
as well as various recent phenomenological models
\cite{Pan:2023hwt,Jakhad:2023mni,Li:2024zze}.
These studies have proposed a variety of assignments for the
observed resonances. Nevertheless, a unified description of the
multiplet structure underlying the three flavor sectors and their
different numbers of experimentally visible states is still
needed. Masses alone are generally insufficient to distinguish
nearby configurations, especially when different states carry
the same total spin and parity. Their strong-decay widths,
dominant decay channels, threshold effects, and finite-charm-mass
mixing must therefore be incorporated into a combined analysis.

In this Letter, we perform a unified study of the $P$-wave
charmed baryons in the $SU(3)$ flavor sextet. We systematically
classify these states within heavy quark effective theory
\cite{Eichten:1989zv,Grinstein:1990mj,Falk:1990yz,
Neubert:1993mb}, and calculate their masses and strong-decay
properties using QCD sum rules
\cite{Shifman:1978bx,Shifman:1978by,Reinders:1984sr,
Colangelo:2000dp,Gubler:2018ctz}
and light-cone sum rules
\cite{Braun:1988qv,Balitsky:1989ry,Chernyak:1990ag,
Ball:1998je}. The same framework and inputs are used for all
three flavor sectors. We explicitly include the mixing between
the two $\lambda$-mode $J^P=3/2^-$ configurations and constrain
it using the observed $\Xi_c$ and $\Omega_c$ spectra. Although
seven $P$-wave states are allowed in each sector, we find that
only four $\Sigma_c$, four $\Xi_c^\prime$, and five $\Omega_c$
states are experimentally resolvable.

As discussed above, the four $\Xi_c$ states and the first four
$\Omega_c$ states listed in Eqs.~\eqref{eq:xic_states} and
\eqref{eq:omega_states} exhibit similar mass-splitting patterns.
Together with our calculated masses and decay properties, these
results lead to their successive assignments as $\lambda$-mode
excitations with $J^P=1/2^-,3/2^-,3/2^-$, and $5/2^-$.
$\Omega_c(3119)$ is tentatively interpreted as a predominantly
$\rho$-mode excitation with $J^P=3/2^-$. The remaining states
are too broad to appear as isolated resonance peaks. The visible
spectrum is therefore not a direct image of the underlying
heavy-quark level scheme: strong decays, threshold effects, and
configuration mixing act as a dynamical filter. The observed
$\Sigma_c(2800)$ and $\Sigma_c(2900)$ structures may contain
four overlapping $P$-wave states, providing concrete
experimental targets for future amplitude analyses of their
spin-parity components.

$\\$
{\it A global picture from heavy quark effective theory}.---
In HQET, a singly charmed baryon consists of a charm quark and
two light quarks. For a $P$-wave excitation, the light-quark
subsystem is characterized by its spin $s_l$ and two possible
orbital configurations. The $\rho$ mode describes an orbital
excitation between the two light quarks, with $l_\rho=1$ and
$l_\lambda=0$, whereas the $\lambda$ mode describes an orbital
excitation between the charm quark and the center of mass of the
two light quarks, with $l_\rho=0$ and $l_\lambda=1$. The color wave function of the two light quarks belongs to the
antisymmetric $\mathbf{\bar 3}_C$ representation. The Pauli
principle therefore requires their combined flavor, spin, and
orbital wave function to be symmetric. For the symmetric
flavor-sextet representation $\mathbf 6_F$, the $\lambda$-mode
orbital wave function is symmetric under the exchange of the two
light quarks and hence requires $s_l=1$, whereas the
antisymmetric $\rho$-mode orbital wave function requires
$s_l=0$, as illustrated in Fig.~\ref{fig:classification}.

We denote these HQET multiplets by
$[\mathbf 6_F(J^P),j_l,s_l,\rho/\lambda]$, where
\begin{equation}
j_l=l_\rho\otimes l_\lambda\otimes s_l,
\qquad
J=j_l\otimes s_c,
\qquad
s_c=1/2.
\label{eq:angular_momenta}
\end{equation}
The flavor-sextet $P$-wave charmed baryons are therefore
classified into four HQET multiplets:
\begin{align}
[\mathbf 6_F,0,1,\lambda]&:
&J^P&=1/2^-,
\nonumber\\
[\mathbf 6_F,1,1,\lambda]&:
&J^P&=\left(1/2^-,3/2^-\right),
\nonumber\\
[\mathbf 6_F,2,1,\lambda]&:
&J^P&=\left(3/2^-,5/2^-\right),
\nonumber\\
[\mathbf 6_F,1,0,\rho]&:
&J^P&=\left(1/2^-,3/2^-\right).
\label{eq:classification}
\end{align}
Thus, HQET predicts seven $P$-wave states in each of the
$\Sigma_c$, $\Xi_c^\prime$, and $\Omega_c$ sectors. The same
multiplet structure underlies all three flavor sectors, whereas
their experimentally visible spectra can differ substantially
because of their distinct strong-decay properties and kinematic
thresholds. We investigate these dynamical effects in the
following sections.

$\\$
{\it Mass spectra, decay properties, and mixing}.---
We study the masses and strong-decay properties of all seven
$P$-wave states in each of the $\Sigma_c$, $\Xi_c^\prime$, and
$\Omega_c$ sectors using QCD sum rules for the masses and
light-cone sum rules for the strong decays within HQET. The
technical framework follows our previous studies of orbitally
excited heavy baryons
\cite{Chen:2015kpa,Chen:2017sci,Yang:2020zjl,Yang:2021lce}.

For a state belonging to the multiplet
$\mathcal{B}=[\mathbf 6_F,j_l,s_l,\rho/\lambda]$, its mass,
including corrections of order $1/m_c$, is written as
\begin{equation}
m_{\mathcal{B},J}
=
m_c+\bar\Lambda_\mathcal{B}
-\frac{1}{2m_c}
\left(
K_\mathcal{B}
+d_{J,j_l}C_{\rm mag}\Sigma_\mathcal{B}
\right),
\label{eq:hqet_mass}
\end{equation}
where $\bar\Lambda_\mathcal{B}$ is the residual energy in the
heavy-quark limit, $K_\mathcal{B}$ and $\Sigma_\mathcal{B}$
denote the kinetic and chromomagnetic corrections, respectively,
$C_{\rm mag}$ is the Wilson coefficient of the chromomagnetic
operator, and $d_{J,j_l}$ is the spin-dependent coefficient.
These quantities are extracted from two-point correlation
functions after Borel transformation and continuum subtraction.

The absolute masses obtained from QCD sum rules generally carry
sizable uncertainties associated with the charm-quark mass,
continuum threshold, Borel parameter, and vacuum condensates.
However, the mass splittings between members of the same
heavy-quark multiplet are significantly more stable, because a
large part of the common theoretical uncertainty cancels. The
multiplet ordering and internal mass splittings therefore provide
particularly useful information for comparison with the parallel
$\Xi_c$ and $\Omega_c$ spectra displayed in
Eq.~\eqref{eq:intro_splittings}.

We calculate the strong couplings of the $P$-wave states to the
ground-state charmed baryons and light pseudoscalar mesons using
light-cone sum rules at leading order in the heavy-quark
expansion. The partial widths are obtained from the corresponding
decay amplitudes and summed over all kinematically allowed
channels. For experimentally established states, we use their
measured masses to evaluate the phase-space factors; for the
predicted $\Sigma_c$ states and the remaining unobserved states,
we use the QCD-sum-rule mass values. The calculated total widths,
dominant decay channels, and partial-wave structures provide
information complementary to the masses and are essential for
determining which states can appear as experimentally resolvable
resonances.

At finite charm-quark mass, the angular momentum $j_l$ of the
light degrees of freedom is no longer exactly conserved, allowing
states with the same flavor and $J^P$ to mix. Of particular
importance is the mixing between the $J^P=3/2^-$ members of the
$[\mathbf 6_F,1,1,\lambda]$ and
$[\mathbf 6_F,2,1,\lambda]$ doublets. We define the physical
states as
\begin{equation}
\begin{pmatrix}
|3/2^-\rangle_L\\
|3/2^-\rangle_H
\end{pmatrix}
=
\begin{pmatrix}
\cos\theta & \sin\theta\\
-\sin\theta & \cos\theta
\end{pmatrix}
\begin{pmatrix}
|[\mathbf 6_F(3/2^-),1,1,\lambda]\rangle\\
|[\mathbf 6_F(3/2^-),2,1,\lambda]\rangle
\end{pmatrix},
\label{eq:mixing}
\end{equation}
where the subscripts $L$ and $H$ denote the lower- and
higher-mass physical states, respectively.

We determine a common mixing angle from the masses and decay
properties of $\Xi_c(2923)$, $\Xi_c(2939)$,
$\Omega_c(3050)$, and $\Omega_c(3066)$, and obtain
\begin{equation}
\theta=37^\circ\pm5^\circ,
\label{eq:common_angle}
\end{equation}
where the quoted $5^\circ$ uncertainty is conservatively assigned
to represent the acceptable variation around the preferred
value. The same mixing pattern is then applied to the
corresponding $J^P=3/2^-$ $\Sigma_c$ states. Small admixtures
among other configurations with the same flavor and $J^P$,
parameterized by the small mixing angles $\theta^\prime$ and
$\theta^{\prime\prime}$ in Table~\ref{tab:result}, are also
allowed at finite charm-quark mass. Although they have little
influence on the mass spectra and total widths, they can generate
otherwise vanishing or strongly suppressed decay amplitudes,
thereby reconciling the dominant configuration assignments with
the experimentally observed channels.

The resulting masses, total widths, dominant decay modes, and
mixing parameters are summarized in Table~\ref{tab:result}. We
next confront these results with the observed $\Xi_c$ and
$\Omega_c$ spectra and use the common multiplet structure thereby
established to reconstruct the remaining broad states,
particularly those in the $\Sigma_c$ sector.

\begin{table*}[hbtp]
\begin{center}
\caption{
Masses, strong-decay properties, and possible experimental
assignments of the $P$-wave charmed baryons in the $SU(3)$
flavor-sextet representation. The broad $\rho$-mode $\Sigma_c$
and $\Xi_c^\prime$ states are omitted for compactness, and only
non-negligible partial widths are quoted. $\Gamma_S$ and
$\Gamma_D$ denote the $S$- and $D$-wave partial decay widths,
respectively. Mixing between the $J^P=3/2^-$ members of the
$[\mathbf 6_F,1,1,\lambda]$ and
$[\mathbf 6_F,2,1,\lambda]$ doublets is included with
$\theta=(37\pm5)^\circ$. The small angles $\theta^\prime$ and
$\theta^{\prime\prime}$ parameterize subleading mixing among
other configurations with the same $J^P$. Entries marked by
$\neq0$ denote channels opened by such mixing; their partial
widths are nonzero but negligible and are therefore omitted from
the quoted total widths. Relevant central masses are quoted to
the nearest MeV to facilitate comparison and display the
splittings, without implying such precision for the absolute
masses.
}
\renewcommand{\arraystretch}{1.5}
\begin{tabular}{   c|c | c | c | c | c | c | c}
\hline\hline
  \multirow{2}{*}{HQET state}&\multirow{2}{*}{Mixing}&\multirow{2}{*}{Mixed state} & Mass & Splitting & Dominant decay modes & Width  & \multirow{2}{*}{Candidate}
\\  &&&   ({GeV}) & ({MeV}) & ({MeV})& ({MeV}) &
\\ \hline\hline
$[\Sigma_c({1\over2}^-),0,1,\lambda]$&\multirow{3}{*}{$\theta^\prime\approx 0^\circ$}&$[\Sigma_c({1\over2}^-),0,1,\lambda]$&$2.830^{+0.060}_{-0.040}$& -- &
$\begin{array}{l}
\Gamma_S\left(\Sigma_c({1/2}^-)\to\Lambda_c \pi\right)=610^{+860}_{-410}
\end{array}$&$610^{+860}_{-410}$&--
\\ \cline{1-1}\cline{3-8}
$[\Sigma_c({1\over2}^-),1,1,\lambda]$&&$[\Sigma_c({1\over2}^-),1,1,\lambda]$&$2.832^{+0.172}_{-0.172}$&\multirow{4}{*}{$30^{+15}_{-21}$}&
$\begin{array}{l}
\Gamma_S\left(\Sigma_c({1/2}^-)\to\Lambda_c \pi\right) \neq 0 \\
\Gamma_S\left(\Sigma_c({1/2}^-)\to\Sigma_c\pi\right)=37^{+58}_{-27}\\
\end{array}$&$37^{+58}_{-27}$ & \multirow{6}{*}{$\Sigma_c(2800)$}
\\ \cline{1-4} \cline{6-7}
$[\Sigma_c({3\over2}^-),1,1,\lambda]$&\multirow{3}{*}{$\theta={37\pm5^\circ}$}&$|\Sigma_c({3\over2}^-)\rangle_L$&$2.862^{+0.167}_{-0.167}$&&
$\begin{array}{l}
\Gamma_D\left(\Sigma_c({3/2}^-)\to\Lambda_c\pi\right)=20^{+65}_{-20}\\
\Gamma_D\left(\Sigma_c({3/2}^-)\to\Sigma_c\pi\right)=7.7^{+37.8}_{-~7.7}\\
\Gamma_S\left(\Sigma_c({3/2}^-)\to\Sigma_c^{*}\pi\right)=7.8^{+12.9}_{-~7.6}
\end{array}$&$35^{+116}_{-~35}$&\multirow{1}{*}{}
\\ \cline{1-1} \cline{3-7}
$[\Sigma_c({3\over2}^-),2,1,\lambda]$&&$|\Sigma_c({3\over2}^-)\rangle_H$&$2.879^{+0.228}_{-0.228}$&\multirow{4}{*}{$68^{+32}_{-35}$}&
$\begin{array}{l}
\Gamma_D\left(\Sigma_c({3/2}^-)\to\Lambda_c\pi\right)=42^{+177}_{-~42}\\
\Gamma_S\left(\Sigma_c({3/2}^-)\to\Sigma_c^{*}\pi\right)=4.4^{+6.3}_{-4.4}\\
\end{array}$&$46^{+183}_{-~46}$&\multirow{2}{*}{$\Sigma_c(2900)$}
\\ \cline{1-4} \cline{6-7}
$[\Sigma_c({5\over2}^-),2,1,\lambda]$&--&$[\Sigma_c({5\over2}^-),2,1,\lambda]$&$2.947^{+0.217}_{-0.217}$&&
$\begin{array}{l}
\Gamma_D\left(\Sigma_c({5/2}^-)\to\Lambda_c\pi\right)=27^{+58}_{-20}\\
\Gamma_D\left(\Sigma_c({5/2}^-)\to \Sigma_c\pi\right)=1.5^{+3.8}_{-1.3}\\
\Gamma_D\left(\Sigma_c({5/2}^-)\to\Sigma_c^{*}\pi\right)=0.9^{+2.2}_{-0.8}
\end{array}$&$29^{+64}_{-22}$&\multirow{1}{*}{}
\\ \hline
$[\Xi_c^\prime({1\over2}^-),0,1,\lambda]$&\multirow{4}{*}{$\theta^\prime \approx 0^\circ$}&$[\Xi_c^{\prime}({1\over2}^-),0,1,\lambda]$&$2.900^{+0.130}_{-0.120}$& -- &
$\begin{array}{l}
\Gamma_S\left(\Xi_c^{\prime}({1/2}^-)\to\Lambda_c \bar K\right)=400^{+610}_{-270}\\
\Gamma_S\left(\Xi_c^{\prime}({1/2}^-)\to \Xi_c \pi\right)=360^{+550}_{-250}
\end{array}$
&$760^{+820}_{-370}$&--
\\ \cline{1-1}\cline{3-8}
$[\Xi_c^\prime({1\over2}^-),1,1,\lambda]$&&$[\Xi_c^\prime({1\over2}^-),1,1,\lambda]$&$2.890^{+0.132}_{-0.132}$&\multirow{5}{*}{$35^{+15}_{-17}$}&
$\begin{array}{l}
\Gamma_S\left(\Xi_c^{\prime}({1/2}^-)\to\Lambda_c \bar K\right) \neq 0 \\
\Gamma_S\left(\Xi_c^{\prime}({1/2}^-)\to \Xi_c \pi\right)  \neq 0 \\
\Gamma_S\left(\Xi_c^{\prime}({1/2}^-)\to \Xi_c^{\prime}\pi\right)=10^{+14}_{-~7}\\
\end{array}$&$10^{+14}_{-~7}$&$\Xi_c(2882)$
\\ \cline{1-4}\cline{6-8}
$[\Xi_c^{\prime}({3\over2}^-),1,1,\lambda]$&\multirow{4}{*}{$\theta={37\pm5^\circ}$}&$|\Xi_c^\prime({3\over2}^-)\rangle_L$&$2.925^{+0.128}_{-0.128}$&&
$\begin{array}{l}
\Gamma_D\left(\Xi_c^{\prime}({3/2}^-)\to \Lambda_c \bar K\right)=1.8^{+2.7}_{-1.1}\\
\Gamma_D\left(\Xi_c^{\prime}({3/2}^-)\to \Xi_c\pi\right)=4.0^{+8.6}_{-3.0}\\
\Gamma_D\left(\Xi_c^{\prime}({3/2}^-)\to\Xi_c^{\prime}\pi\right)=1.6^{+2.0}_{-1.0}\\
\Gamma_S\left(\Xi_c^{\prime}({3/2}^-)\to \Xi_c^{*}\pi\right)=2.0^{+2.9}_{-1.5}
\end{array}$&$9^{+16}_{-~7}$
& $\Xi_c(2923)$
\\ \cline{1-1}\cline{3-8}
$[\Xi_c^\prime({3\over2}^-),2,1,\lambda]$&&$|\Xi_c^\prime({3\over2}^-)\rangle_H$&$2.936^{+0.238}_{-0.238}$&\multirow{4}{*}{$64^{+30}_{-31}$}&
$\begin{array}{l}
\Gamma_D\left(\Xi_c^{\prime}({3/2}^-)\to \Lambda_c \bar K\right)=4.3^{+6.3}_{-2.6}\\
\Gamma_D\left(\Xi_c^{\prime}({3/2}^-)\to \Xi_c\pi\right)=9^{+18}_{-~6}\\
\Gamma_S\left(\Xi_c^{\prime}({3/2}^-)\to \Xi_c^{*}\pi\right)= 1.2^{+1.8}_{-0.9}
\end{array}$&$15^{+26}_{-10}$&$\Xi_c(2939)$
\\ \cline{1-4} \cline{6-8}
$[\Xi_c^\prime({5\over2}^-),2,1,\lambda]$&--&$[\Xi_c^\prime({5\over2}^-),2,1,\lambda]$&$3.000^{+0.226}_{-0.226}$&&
$\begin{array}{l}
\Gamma_D\left(\Xi_c^{\prime}({5/2}^-)\to \Lambda_c \bar K\right)=3.2^{+7.1}_{-2.4}\\
\Gamma_D\left(\Xi_c^{\prime}({5/2}^-)\to \Xi_c \pi\right)=5.9^{+8.1}_{-3.4}\\
\Gamma_D\left(\Xi_c^{\prime}({5/2}^-)\to \Xi_c^{*} \pi\right)=0.16^{+0.35}_{-0.13}
\end{array}$&$9^{+15}_{-~6}$
&$\Xi_c(2965)$
\\ \hline
$[\Omega_c({1\over2}^-),0,1,\lambda]$&\multirow{2}{*}{$\theta^\prime \approx0^\circ$}&$[\Omega_c({1\over2}^-),0,1,\lambda]$&$3.030^{+0.180}_{-0.190}$& -- &$\begin{array}{l}
\Gamma_S\left(\Omega_c({1/2}^-)\to\Xi_c \bar K\right)=980^{+1530}_{-~670}
\end{array}$&$980^{+1530}_{-~670}$&--
\\ \cline{1-1} \cline{3-8}
$[\Omega_c({1\over2}^-),1,1,\lambda]$&&$[\Omega_c({1\over2}^-),1,1,\lambda]$&$3.019^{+0.106}_{-0.106}$& \multirow{2}{*}{$31^{+14}_{-18}$} & $\Gamma_S\left(\Omega_c({1/2}^-)\to\Xi_c \bar K\right) \neq 0$ &$\sim~0$&$\Omega_c(3000)$
\\ \cline{1-4} \cline{6-8}
$[\Omega_c({3\over2}^-),1,1,\lambda]$&\multirow{2}{*}{$\theta=37\pm5^\circ$}&$|\Omega_c({3\over2}^-)\rangle_L$&$3.050^{+0.102}_{-0.102}$&  &
$\begin{array}{l}
\Gamma_D\left(\Omega_c({3/2}^-)\to \Xi_c \bar K\right)=1.3^{+2.6}_{-1.0}
\end{array}$&$1.3^{+2.6}_{-1.0}$&$\Omega_c(3050)$
\\ \cline{1-1} \cline{3-8}
$[\Omega_c({3\over2}^-),2,1,\lambda]$&&$|\Omega_c({3\over2}^-)\rangle_H$&$3.064^{+0.234}_{-0.234}$&\multirow{2}{*}{$55^{+27}_{-29}$}&$\begin{array}{l}
\Gamma_D\left(\Omega_c({3/2}^-)\to \Xi_c \bar K\right)=3.7^{+7.3}_{-2.9}
\end{array}$&$3.7^{+7.3}_{-2.9}$&$\Omega_c(3066)$
\\ \cline{1-4} \cline{6-8}
$[\Omega_c({5\over2}^-),2,1,\lambda]$&--&$[\Omega_c({5\over2}^-),2,1,\lambda]$&$3.119^{+0.217}_{-0.217}$&&
$\begin{array}{l}
\Gamma_D\left(\Omega_c({5/2}^-)\to \Xi_c \bar K\right)=3.3^{+6.4}_{-2.5}
\end{array}$&$3.3^{+6.4}_{-2.5}$&$\Omega_c(3090)$
\\ \hline
\multirow{1}{*}{$[\Omega_c({1\over2}^-),1,0,\rho]$}&\multirow{1}{*}{--}&\multirow{1}{*}{$[\Omega_c({1\over2}^-),1,0,\rho]$}&\multirow{1}{*}{$3.106^{+0.159}_{-0.159}$}& \multirow{3}{*}{$11^{+5}_{-5}$} &$\Gamma_S\left(\Omega_c({1/2}^-)\to\Xi_c^\prime \bar K\right) =260^{+540}_{-260}$ &\multirow{1}{*}{$260^{+540}_{-260}$}&\multirow{1}{*}{--}
\\ \cline{1-4} \cline{6-8}
\multirow{2}{*}{$[\Omega_c({3\over2}^-),1,0,\rho]$}&\multirow{2}{*}{$\theta^{\prime\prime} \approx0^\circ$}&\multirow{2}{*}{$[\Omega_c({3\over2}^-),1,0,\rho]$}&\multirow{2}{*}{$3.117^{+0.160}_{-0.160}$}&& $\Gamma_D\left(\Omega_c({3/2}^-)\to\Xi_c \bar K\right) \neq 0$ &\multirow{2}{*}{$0.3^{+0.5}_{-0.2}$}
&\multirow{2}{*}{$\Omega_c(3119)$}
\\ 
&&&&&$\Gamma_D\left(\Omega_c({3/2}^-)\to\Xi_c^\prime \bar K\right) =0.3^{+0.5}_{-0.2}$&
\\ \hline\hline
\end{tabular}
\label{tab:result}
\end{center}
\end{table*}

$\\$
{\it A unified interpretation of the observed spectra}.---
Using the level ordering in Eq.~\eqref{eq:intro_splittings},
together with our calculated masses and decay properties, we
obtain the parallel assignments
\begin{align}
\Xi_c(2882),\ \Omega_c(3000)
&\longleftrightarrow
[\mathbf 6_F(1/2^-),1,1,\lambda],
\nonumber\\
\Xi_c(2923),\ \Omega_c(3050)
&\longleftrightarrow
|3/2^-\rangle_L,
\nonumber\\
\Xi_c(2939),\ \Omega_c(3066)
&\longleftrightarrow
|3/2^-\rangle_H,
\nonumber\\
\Xi_c(2965),\ \Omega_c(3090)
&\longleftrightarrow
[\mathbf 6_F(5/2^-),2,1,\lambda].
\label{eq:physical_assignments}
\end{align}
Here, $|3/2^-\rangle_L$ and $|3/2^-\rangle_H$ denote the
lower- and higher-mass mixtures defined in
Eq.~\eqref{eq:mixing}. The parallel ordering of the observed
$\Xi_c$ and $\Omega_c$ states supports a common $\lambda$-mode
multiplet structure. The calculated strong-decay properties
provide further support: the predicted total widths are
consistent with experiment, while the dominant decay modes agree
with the observed channels.

The fifth narrow $\Omega_c$ state is tentatively identified as
\begin{equation}
\Omega_c(3119)
\longleftrightarrow
[\Omega_c(3/2^-),1,0,\rho],
\label{eq:omega3119_assignment}
\end{equation}
up to small admixtures from other $J^P=3/2^-$ configurations.
Its narrow width arises because the $S$-wave
$\Xi_c^*\bar K$ channel is closed, whereas its
$J^P=1/2^-$ partner is broad owing to the open $S$-wave
$\Xi_c^\prime\bar K$ channel. A small finite-charm-mass
admixture generates the otherwise vanishing $\Xi_c\bar K$
amplitude, allowing $\Omega_c(3119)$ to be observed in this
channel while remaining narrow. The same decay dynamics renders
the remaining three $\Xi_c^\prime$ states and the
$[\Omega_c(1/2^-),0,1,\lambda]$ and
$[\Omega_c(1/2^-),1,0,\rho]$ states broad through open
strong-decay channels. Their absence as narrow peaks is therefore
consistent with the complete seven-state HQET classification.

Having established a common multiplet ordering and mixing pattern
from the narrow $\Xi_c$ and $\Omega_c$ spectra, we apply the same
structure to the $\Sigma_c$ sector. We predict four relatively
resolvable $P$-wave $\Sigma_c$ states corresponding to the four
$\lambda$-mode levels in Eq.~\eqref{eq:physical_assignments},
while the remaining three states are substantially broader. The
observed $\Sigma_c(2800)$ and $\Sigma_c(2900)$ structures may
therefore contain several overlapping $P$-wave components rather
than correspond one-to-one to individual states. Future amplitude
analyses of the $\Lambda_c\pi$ spectrum may resolve their
spin-parity and partial-wave contributions.

More generally, the narrow $\Xi_c$ and $\Omega_c$ states provide spectroscopic anchors for reconstructing the remaining broad states in the same HQET multiplets. A comprehensive extension to the corresponding $P$-wave flavor-sextet bottom baryons, together with detailed decay analyses and predictions for missing states, is presented in a companion study~\cite{Zhang:2026ugd}. The visible spectrum is not a direct image of the underlying HQET level scheme, but is a dynamically filtered manifestation shaped by strong decays, kinematic thresholds, and configuration mixing.

\begin{acknowledgements}
This project is supported by
the National Natural Science Foundation of China under Grant No.~12075019,
the Jiangsu Provincial Double-Innovation Program under Grant No.~JSSCRC2021488,
and
the Fundamental Research Funds for the Central Universities.
\end{acknowledgements}

\bibliographystyle{elsarticle-num}
\bibliography{ref}

\end{document}